\documentclass[twocolumn]{aastex702}

\usepackage{amsmath,amssymb}
\usepackage{booktabs}
\usepackage{subcaption}
\usepackage{enumitem}

\begin{document}

\title{Constraining the Fraction of Little Red Dots in JADES Protoclusters at $3 \leq z < 6$}

\correspondingauthor{Carl Audric Guia}
\email{cgguia@college.harvard.edu}

\author[0009-0009-7258-1637]{Carl Audric Guia}
\affiliation{Center for Astrophysics $|$ Harvard \& Smithsonian, 60 Garden Street, Cambridge, MA 02138, USA}
\email{cgguia@college.harvard.edu}

\author[0000-0002-4622-6617]{Fengwu Sun}
\affiliation{Center for Astrophysics $|$ Harvard \& Smithsonian, 60 Garden Street, Cambridge, MA 02138, USA}
\email{fengwu.sun@cfa.harvard.edu}

\author[0000-0002-2929-3121]{Daniel J. Eisenstein}
\affiliation{Center for Astrophysics $|$ Harvard \& Smithsonian, 60 Garden Street, Cambridge, MA 02138, USA}
\email{deisenstein@cfa.harvard.edu}

\author[0000-0002-4271-0364]{Brant E. Robertson}
\affiliation{Department of Astronomy \& Astrophysics, University of California Santa Cruz, 1156 High St, Santa Cruz, CA 95064, USA}
\affiliation{Institute of Astronomy, University of Cambridge, Madingley Road, Cambridge CB3 0HA, UK}
\email{brant@ucsc.edu}

\author[0000-0002-8651-9879]{Andrew J. Bunker}
\affiliation{Department of Physics, University of Oxford, Denys Wilkinson Building, Keble Road, Oxford OX1 3RH, UK}
\email{andy.bunker@physics.ox.ac.uk}

\begin{abstract}
We measure the fraction of little red dots (LRDs) among H$\alpha$-emitting galaxies in protocluster environments, extending environmental studies of LRDs below $z=4$. Using JADES NIRCam F322W2 $+$ F356W $+$ F444W grism data from the GTO-4540 program, we probe down to $z \approx 2.7$, below the $z \approx 3.75$ limit of the F356W$+$F444W surveys previously used to measure the LRD fraction from grism spectroscopy. The survey of 1,445 spectroscopically confirmed galaxies at $z=0-9$ enable the identification of 16 protocluster candidates via density-based clustering and one LRD via its broad H$\alpha$ emission and compact morphology. We extend the sample with the deeper NIRSpec classifications of the JADES ``Dark Horse'' program, yielding eight LRDs at $3.4 \le z \le 6.6$ and two broad-line active galactic nuclei (BLAGN) that are not LRDs. The protocluster LRD fraction is low across the full redshift range with no significant trend once luminosity-matched. In every redshift bin, the LRD fractions in the protocluster and field populations are statistically indistinguishable, showing no evidence for a  dependence on large-scale environment. This is consistent with---and extends to $3 \le z < 4$---the environment-independent AGN fraction reported at $3.9 < z < 6$. We discuss the implications of an environment-independent LRD fraction for early black-hole growth and for the fate of high-redshift LRDs in the lower-redshift protoclusters that descend from them.
\end{abstract}

\section{Introduction} \label{sec:introduction}

Among the surprising discoveries of JWST is a population of little red dots (LRDs) \citep{2023ApJ...954L...4K, 2023Natur.616..266L, 2024A&A...691A.145M, 2023ApJ...959...39H, 2024ApJ...963..129M}. They are strikingly red, compact, and observed across a wide range of redshifts. \citet{2025ApJ...986..126K} found that LRDs are abundant at high redshift ($z\sim4-8$) but rare at lower redshift ($z<4$).  Because instruments covering the same rest-frame wavelengths can achieve higher sensitivities at lower redshift, the apparent sharp drop in the population of LRDs at $z\sim4$ has been interpreted as genuine evolution of the population rather than an observational effect \citep{2025ApJ...986..126K}.

LRDs have unusual properties. They are extremely compact \citep{2023ApJ...955L..12B, 2023ApJ...952..142F} and reach immense densities when modeled as purely stellar systems \citep{2024RNAAS...8..207G}. At the same time, many show broad Balmer emission lines, usually taken as a signature of active galactic nuclei (AGN) \citep{2024ApJ...964...39G}, together with a distinctive V-shaped spectral energy distribution (SED) whose origin remains debated \citep{2026ApJ..1006..205R}. Yet unlike typical AGN, LRDs are very X-ray weak \citep{2025ApJ...986..126K, 2026ApJ...997..364L, 2024ApJ...974L..26Y, 2025MNRAS.538.1921M}. They are also radio-quiet in deep stacking, which disfavors their interpretation as scaled-down super-Eddington quasars \citep{2025RNAAS...9..355G}. These properties are difficult to reconcile within standard galaxy or AGN models.

The demographics of LRDs add a crucial clue that any model must match. \citet{2025ApJ...986..126K} present a sample of 341 LRDs at $z\sim2-11$ whose numbers grow with cosmic time from $z\sim8$ and then drop sharply below $z\sim4$. Whether this decline is physical or a selection effect remains debated: \citet{2026arXiv260630253L} find no significant evolution in LRD abundance at $z>2$ and attribute the apparent low-redshift drop to sampling bias, while \citet{2026arXiv260407138R} show that the inferred evolution depends strongly on selection strategy. At the lowest redshifts, \citet{2026ApJ...997..364L} report three LRDs at $z=0.1-0.2$, with further low-redshift examples since identified in DESI \citep[e.g.,][]{2026A&A...709L..11D, 2026arXiv260514233P}. Recent grism and imaging surveys like NEXUS \citep{2026ApJ...999...31Z} and EIGER \citep{2026arXiv260700084K} have begun to identify and characterize this population below $z\sim4$. However, its dependence on environment at these redshifts remains unmeasured. 

Protoclusters, the overdense progenitors of present-day galaxy clusters, offer an ideal setting to study LRDs. Protoclusters trace the densest environments above $z\sim1.5$, the epoch before most clusters virialized \citep{2015MNRAS.452.2528M}. Protoclusters have been predicted at high redshifts $z\sim3-5$ \citep{1972A&A....20..189S} and have now been identified out to $z>8$  \citep{2016A&ARv..24...14O, 2025arXiv251011770F}. They are expected to host a substantial share of star formation in the universe: up to 20\% at $z=2$ and 50\% at $z=10$ \citep{2017ApJ...844L..23C}. Their galaxies also undergo accelerated assembly \citep{2013ApJ...773..154L}, so LRDs may form earlier or differently in these environments.

At $3.9<z<6$, \citet{2026ApJ...997...61L} defined the AGN fraction as the proportion of broad-line AGN (BLAGN) and LRDs among H$\alpha$ emitters (HAEs). They measured its dependence on large-scale overdensity, quantified as the HAE overdensity within a cylinder of radius $8.5\,h^{-1}$ and length $15\,h^{-1}$\,cMpc around each source. Since HAEs are tracers of the general star-forming population, comparing the abundance of LRD$+$BLAGN to HAEs tests whether black-hole accretion is enhanced relative to normal star formation in dense environments like protoclusters. In the Great Observatories Origins Deep Survey North (GOODS-N) field, \citet{2026ApJ...997...61L} found average AGN fractions of 3.7\% at $3.9<z<5$, 4.0\% at $5<z<6$, and 3.9\% over $3.9<z<6$, with no dependence on large-scale overdensity. This result suggests that the growth of LRD$+$BLAGN is not strongly regulated by large-scale environment at these redshifts. 

Whether this environmental independence persists below $z=3.9$ is untested. In this work, we extend the analysis of \citet{2026ApJ...997...61L} to lower redshift by measuring the fraction of LRDs among HAEs in protoclusters and field environments, together with the combined LRD$+$BLAGN fraction for direct comparison with \citet{2026ApJ...997...61L}. Because protoclusters at $3\le z<4$ are the descendants of those at higher redshift, the LRD fraction in these more evolved protoclusters constrains the fate of the high-redshift LRD population. We address two questions:
\begin{enumerate}[itemsep=0pt, topsep=0pt, parsep=0pt]
    \item Environment: If the protocluster LRD fraction at $z<3.9$ matches the field fraction at the same redshift, the LRD fraction is independent of environment, extending the result of \citet{2026ApJ...997...61L} to lower redshift.
    \item Evolution: If the protocluster LRD fraction is lower than at higher redshift while the LRD$+$BLAGN fraction is unchanged, LRDs may transform into normal BLAGN near $z\sim4$, the threshold at which \citet{2025ApJ...986..126K} report a rapid decline in the LRD population. If the combined fraction is also lower, LRDs may instead be short-lived sources that fade or merge with other galaxies and AGN.
\end{enumerate}

Following \citet{2026ApJ...997...61L}, we adopt a flat $\Lambda$CDM cosmology with $H_0=70~\mathrm{km\,s^{-1}\,Mpc^{-1}}$, $\Omega_{m,0}=0.3$, and $\Omega_{\Lambda,0}=0.7$.

\section{Data} \label{sec:Data}

 The JWST Advanced Deep Extragalactic Survey (JADES) Origins Field (JOF) is one of the deepest fields observed to date \citep{2025ApJS..281...50E}. Our primary dataset is the Cycle-3 GTO-4540 NIRCam Grism Spectroscopy program (PI: D. J. Eisenstein).  Its $2.4$ to $5 \ \mu$m coverage over most of the JOF captures H$\alpha$ lines at $2.7 < z < 6.6$ across the F322W2, F356W, and F444W filters. The resulting data products include direct NIRCam images that provide morphological information, as well as 1D and 2D spectra that form the basis for identifying HAEs and AGNs. Within its footprint, the grism spectroscopy, aided by the JOF medium-band imaging, detects nearly all HAEs with $F_{\mathrm{H}\alpha}\gtrsim2\times10^{-18}\,\mathrm{erg\,s^{-1}\,cm^{-2}}$, the flux limit we adopt for the wide-field slitless spectroscopy (WFSS). This provides a near-complete census of HAEs above this grism detection limit at $z\sim3-5$.  The resulting catalog contains 1,445 NIRCam grism redshifts at $z\approx0$--9. 

The second dataset is the JADES NIRCam photometric catalog (version 2.0; \citealt{2025ApJS..281...50E}), which covers the deep region of the  GOODS-South (GOODS-S) field \citep{2004ApJ...600L..93G}, including the JOF. It contains photometry for $\sim$94{,}000 sources in 14 NIRCam and 9 HST filters, measured in seven circular apertures, and we use it to quantify the compactness of the LRD candidates (Section~\ref{sec:lrdsample}). Source IDs throughout follow the JADES photometric catalog used for the GTO-4540 extraction; they match the v2.0 catalog except for ID~634042 (v2.0 ID~39376) and ID~642396, which come from a later catalog version.

A third dataset comes from the JADES ``Dark Horse'' (DH) survey \citep{2026MNRAS.549ag824D}, a dense-shutter NIRSpec/MSA program (PIs: D. J. Eisenstein and R. Maiolino) covering the same JOF. Dark Horse complements the JADES NIRSpec data releases \citep{2024A&A...690A.288B,2025ApJS..277....4D,2026MNRAS.549ag836C,2026MNRAS.549ag939S}. We do not use the redshifts from those releases so that protocluster identification rests on the uniformly selected grism sample. The only NIRSpec redshifts we add are those of the Dark Horse AGN (Section~\ref{sec:lrdsample}). Dark Horse targeted ${\sim}854$ sources and secured robust spectroscopic redshifts for 539 galaxies at $2.5 \lesssim z \lesssim 8.9$, reaching emission-line sensitivities roughly five times deeper than the NIRCam grism near $4\,\mu$m. Its greater depth and spectral resolution resolve broad Balmer lines and V-shaped continua in faint AGN that the grism cannot detect, providing deeper and independent LRD classifications in the same field. \citet{2026MNRAS.549ag824D} classify a broad-line AGN as an LRD when its SED is V-shaped. We use these classifications to extend and validate our grism-selected LRD sample. It is worth noting that Dark Horse is not a uniform census of AGN as it prioritized LRD candidates \citep{2026MNRAS.549ag824D}. Its targeting was based on photometry alone and is blind to the environments we identify, so any resulting incompleteness affects the protocluster and field HAEs equally without biasing their comparison.

\begin{figure*}[htbp]
    \centering
    \includegraphics[width=\textwidth]{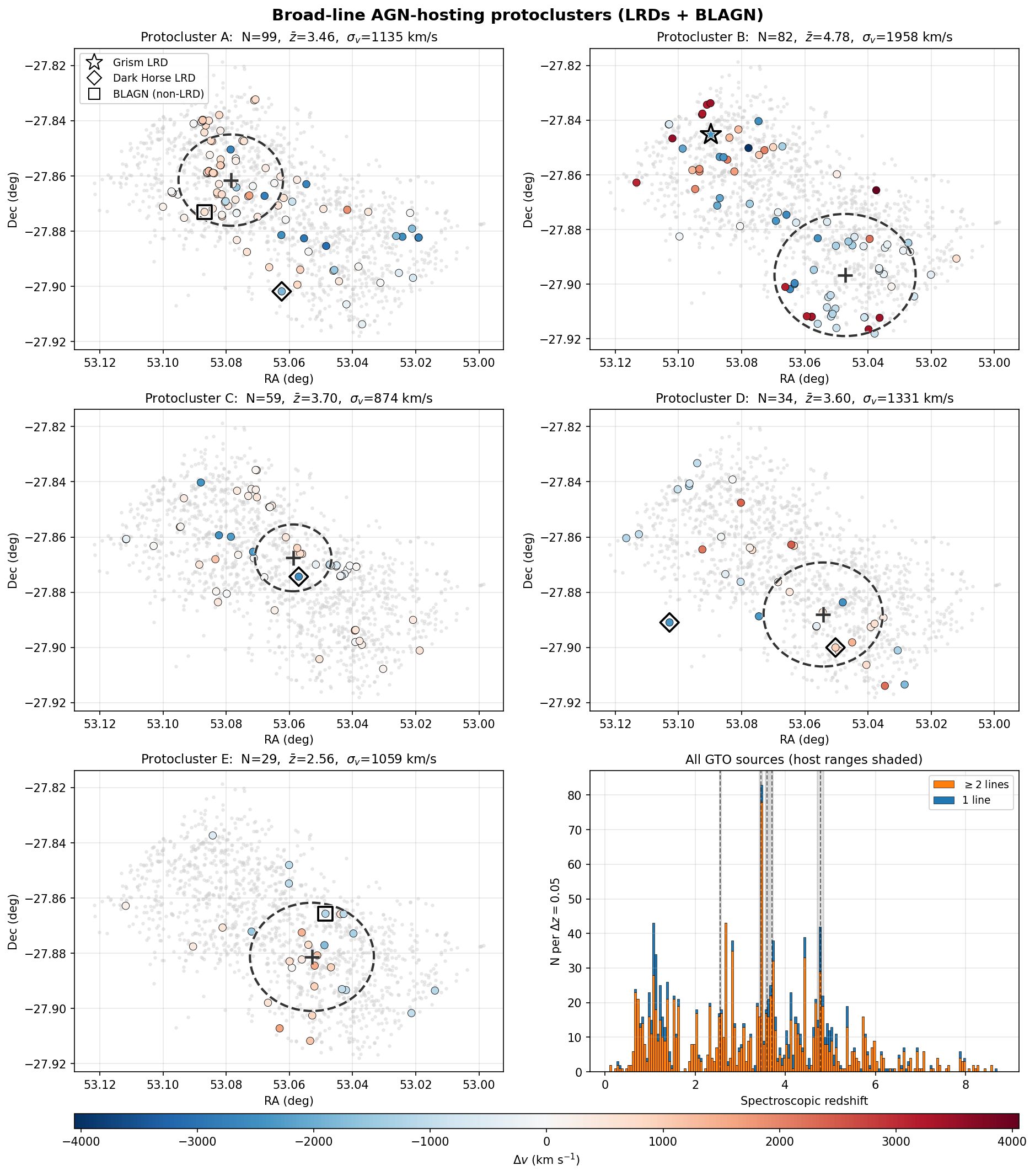}
    \caption{Sky distribution (R.A.--Dec.) of the protocluster candidates that host at least one LRD or BLAGN. Each panel is a single protocluster (labeled A--E); members are colored by their line-of-sight velocity offset $\Delta v$ from the group mean, with the kernel-density center ($+$) and the projected $R_{200}$ (dashed circle) drawn in charcoal. Sources hosting a broad line are ringed by class: grism-selected LRD (star), Dark Horse LRD (diamond), and BLAGN that are not LRDs (square). The remaining panel shows the  spectroscopic-redshift distribution of all GTO-4540 sources, with each host protocluster's redshift range shaded; the stacked histogram separates   sources with $\geq2$ detected emission lines from single-line detections.}
    \label{fig:hosts}
\end{figure*}

\section{Results}

\subsection{Protocluster and Field Samples} \label{sec:samples}

We identified protocluster candidates by applying the density-based algorithm DBSCAN \citep{10.5555/3001460.3001507} to the comoving positions of the spectroscopically confirmed galaxies. The algorithm used anisotropic linking lengths $\ell_\perp=10$ and $\ell_\parallel=20$ cMpc to accommodate redshift-space distortion. With the requirement of at least 20 members per group, the clustering yielded 16 protocluster candidates at $1.1\lesssim z\lesssim5.8$, ten of which lie at $3\le z<6$ (Appendix~\ref{app:clustering}, Table~\ref{tab:cands}). Because the JOF spans only $\approx$8--13\,cMpc transversely, comparable to $\ell_\perp$, these candidates are effectively overdensities along the line of sight; we test their significance in Appendix~\ref{app:clustering}. Our protocluster population comprises the members of all 16 candidates; at $2.5\le z<6$, 340 of the 548 detectable HAEs (62\%) are protocluster members. Galaxies assigned to no group formed the field sample. 

For each candidate, we measured a line-of-sight velocity dispersion $\sigma_v = c\,\sigma_z/(1+\bar z)$, where $\bar z$ and $\sigma_z$ are the mean and standard deviation of the members' spectroscopic redshifts. We also measured a virial mass proxy $M_{200}\propto\sigma_v^3/H(\bar z)$ to derive the corresponding radius estimate $R_{200}$. Five of the 16 candidates host at least one LRD or BLAGN (Figure~\ref{fig:hosts}). They have $\sigma_v\sim870$--$1960\,\mathrm{km\,s^{-1}}$ and $\log(M_{200}/M_\odot)\sim14$--$15$. These dispersions are large for protoclusters and, together with the corresponding $M_{200}$, should be read as order-of-magnitude proxies rather than virial masses. The line-of-sight linking admits a broad redshift window that might inflate $\sigma_v$; candidate B, for example, appears to comprise two structures at $z\approx4.75$ and $4.83$ joined by the linking.  In addition, these still-assembling structures are not yet expected to be virialized at these redshifts.  Figure~\ref{fig:hosts} shows their member galaxies, color-coded by line-of-sight velocity offset, and their $R_{200}$.

\begin{figure*}[htbp]
    \centering
    \includegraphics[width=\textwidth]{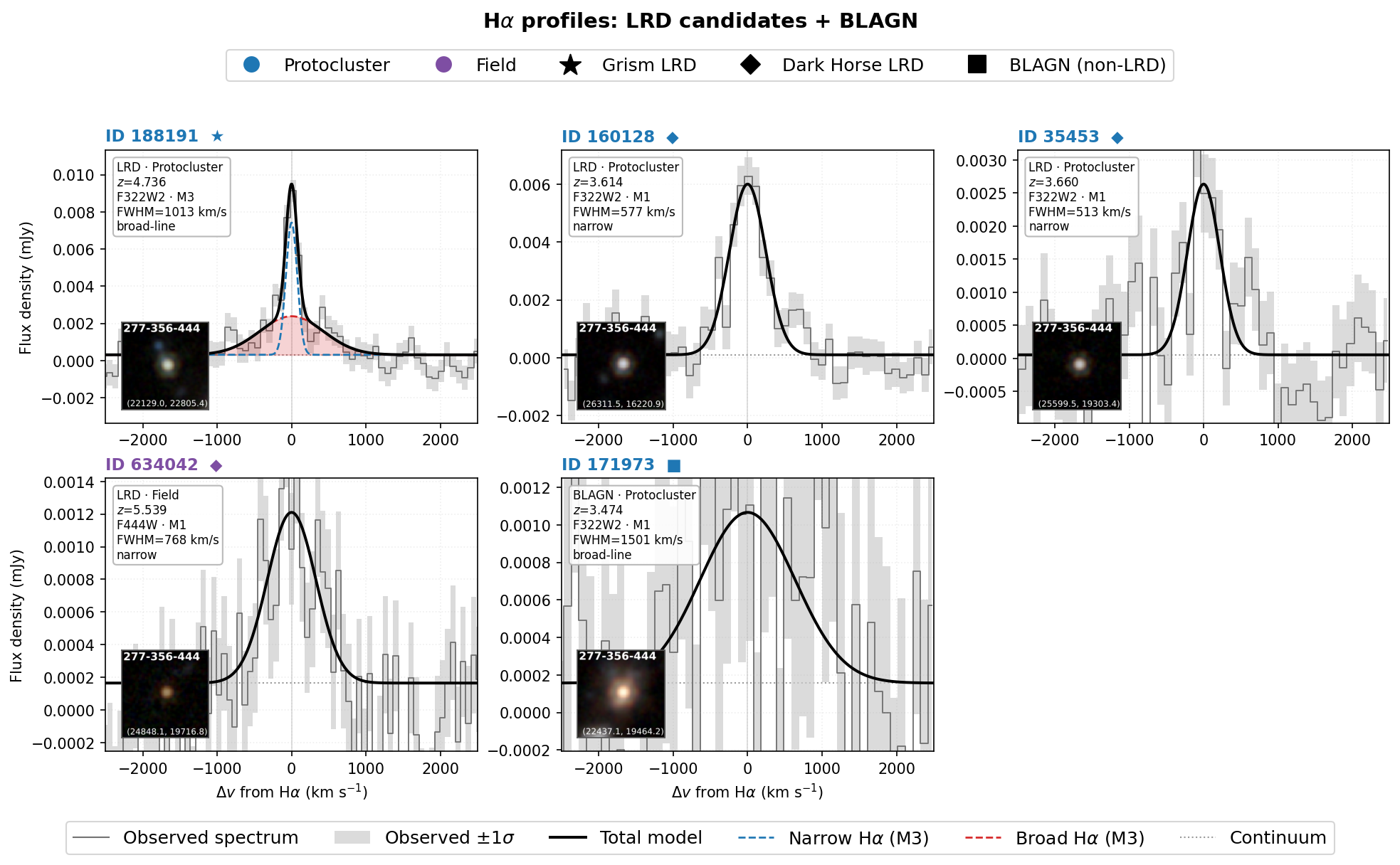}
    \caption{Best-fit H$\alpha$ profiles of the four LRDs and the one BLAGN with a NIRCam WFSS grism H$\alpha$ detection. The observed optimal-extraction spectrum (gray step, with its $\pm1\sigma$ band) is shown with the best-fit model: a single Gaussian (Model~1) or a narrow$+$broad decomposition (Model~3; dashed blue and red), on a constant continuum (dotted). Each panel is centered on H$\alpha$ in velocity space and annotated with redshift, filter, model, and the FWHM of the dominant component. Title color encodes environment (protocluster blue, field purple). Glyphs encode classification as in Figure~\ref{fig:hosts}: grism-selected LRD (star), Dark Horse LRD (diamond), and BLAGN that is not an LRD (square). Insets show $1.9''\times1.9''$ F277W/F356W/F444W NIRCam cutouts, illustrating the compact morphology. The BLAGN ID~171973 is only marginally detected in the grism; its classification rests on the NIRSpec spectrum of \citet{2026MNRAS.549ag824D}.}
    \label{fig:fits}
\end{figure*}

\begin{table*}[htbp]
\centering
\setlength{\tabcolsep}{5pt}
\caption{LRD sample. ``Classified by'' gives the source of the LRD classification: our NIRCam grism (Section~\ref{sec:lrdsample}) or Dark Horse NIRSpec \citep{2026MNRAS.549ag824D}. The grism columns give our fit to the GTO-4540 H$\alpha$ line wherever the grism detects it, including for Dark Horse LRDs. Model~1 is a single Gaussian, and Model~3 is a narrow$+$broad component pair (Appendix~\ref{app:fitting}); FWHM is in km\,s$^{-1}$. The grism fits of the Dark Horse LRDs are narrow; their broad lines are resolved only by NIRSpec. CIRC3/CIRC1 is the F444W compactness ratio (Section~\ref{sec:lrdsample}). Ellipses mark sources without a grism H$\alpha$ detection or a photometric counterpart.}
\label{tab:lrd}
\begin{tabular}{@{}rrllccrcrr@{}}
\toprule
ID & $z$ & Environment & Classified by & \multicolumn{4}{c}{NIRCam grism H$\alpha$ fit} & CIRC3/CIRC1 & $\log L_{\mathrm{H}\alpha}$ \\
\cmidrule(lr){5-8}
   &     &             &         & Filter & Model & FWHM & Broad? &  & (erg\,s$^{-1}$) \\
\midrule
5070   & 3.6437 & Field        & Dark Horse & \nodata & \nodata & \nodata & \nodata & 1.039   & \nodata \\
634042 & 5.5389 & Field        & Dark Horse & F444W   & 1       & 768     & No      & 1.026   & 41.94 \\
642396 & 6.5535 & Field        & Dark Horse & \nodata & \nodata & \nodata & \nodata & \nodata & \nodata \\
5756   & 3.4373 & Protocluster & Dark Horse & \nodata & \nodata & \nodata & \nodata & 1.060   & \nodata \\
13064  & 3.5633 & Protocluster & Dark Horse & \nodata & \nodata & \nodata & \nodata & 1.050   & \nodata \\
160128 & 3.6140 & Protocluster & Dark Horse & F322W2  & 1       & 577     & No      & 1.045   & 42.19 \\
35453  & 3.6598 & Protocluster & Dark Horse & F322W2  & 1       & 513     & No      & 1.008   & 41.78 \\
188191 & 4.7362 & Protocluster & Grism      & F322W2  & 3       & 1013    & Yes     & 1.068   & 42.34 \\
\bottomrule
\end{tabular}
\end{table*}
\begin{table*}[htbp]
\centering
\caption{BLAGN from the Dark Horse sample that are not classified as LRDs. These sources enter only the combined LRD$+$BLAGN fraction; 175773, lacking grism H$\alpha$, enters no fraction.}
\label{tab:blagn}
\begin{tabular}{rrll}
\toprule
ID & $z$ & Environment & Note \\
\midrule
175773 & 2.5447 & Protocluster & tentative broad, BPT type-2; no grism H$\alpha$ ($z=2.55$, blueward of F322W2) \\
171973 & 3.4739 & Protocluster  & X-ray det. \citep{2025ApJ...986..126K}; enters LRD$+$BLAGN (grism H$\alpha$, $z=3.47$) \\
\bottomrule
\end{tabular}
\end{table*}

\subsection{LRD and BLAGN Sample} \label{sec:lrdsample}

We selected LRDs as compact, spatially unresolved sources with broad H$\alpha$ emission. Compactness was quantified by the ratio of F444W fluxes in circular apertures of radius $0.25''$ (CIRC3) and $0.10''$ (CIRC1), each corrected to total using the point spread function, so that unresolved sources have CIRC3/CIRC1 $\approx1$. We require CIRC3/CIRC1 $<1.5$ \citep[after][]{2025ApJ...985..119Z}, confirmed by visual inspection of the NIRCam imaging. Broad H$\alpha$ was identified from Gaussian line fits with FWHM $>1000\,\mathrm{km\,s^{-1}}$; the appropriate model was chosen by the Bayesian information criterion (Appendix~\ref{app:fitting}). The grism data yielded four broad-line candidates. ID~175930 is broad only in a BIC-disfavored filter, and ID~176606 is spatially resolved in the F444W imaging (CIRC3/CIRC1 $=1.33$, against $\le1.07$ for the rest of the sample). We exclude both of them from the sample (Appendix~\ref{app:robustness}). ID~171973 is only marginally detected in the grism (signal-to-noise $\approx4$) and is classified by Dark Horse as a BLAGN that is not an LRD (below). This leaves one grism-selected LRD, ID~188191, which lies in a protocluster.

We extended the sample with LRDs classified from the Dark Horse NIRSpec spectroscopy \citep{2026MNRAS.549ag824D}, adding seven LRDs and two BLAGN that are not LRDs.  The grism misses these LRDs for three reasons. Three (5070, 13064, and 642396) lie outside the GTO-4540 coverage, and one (5756) is too faint in H$\alpha$ for a grism detection. The other three (160128, 35453, and 634042) have grism H$\alpha$ detections that appear narrow (Table~\ref{tab:lrd}); their broad components fall below the grism sensitivity and are resolved only by NIRSpec. To place the Dark Horse sources in our environments, we added the six without a grism redshift to the catalog at their NIRSpec redshifts and re-ran DBSCAN with the same clustering parameters as before, recovering the same 16 protocluster candidates. The final sample comprises eight LRDs at $3.4\le z\le6.6$ and two BLAGN (Tables~\ref{tab:lrd} and \ref{tab:blagn}). Only sources with a grism H$\alpha$ detection count as detectable HAEs, so 5070, 5756, 13064, and 642396 do not enter the fractions. The H$\alpha$ profiles and NIRCam cutouts of the LRDs are shown in Figure~\ref{fig:fits}. The classifications are robust to the compactness aperture (Appendix~\ref{app:robustness}). 

Our grism criterion (compactness and broad H$\alpha$) does not impose the V-shaped SED required by the Dark Horse classifications. It thus cannot distinguish LRDs from compact, blue broad-line AGN (``little blue dots''; \citealt{2026A&A...713A.332B}). This affects 188191, our sole LRD from the grism.  Our environmental comparison is unchanged whether 188191 is treated as an LRD, treated as a BLAGN, or excluded.

\begin{figure*}[htbp]
    \centering
    \includegraphics[width=\textwidth]{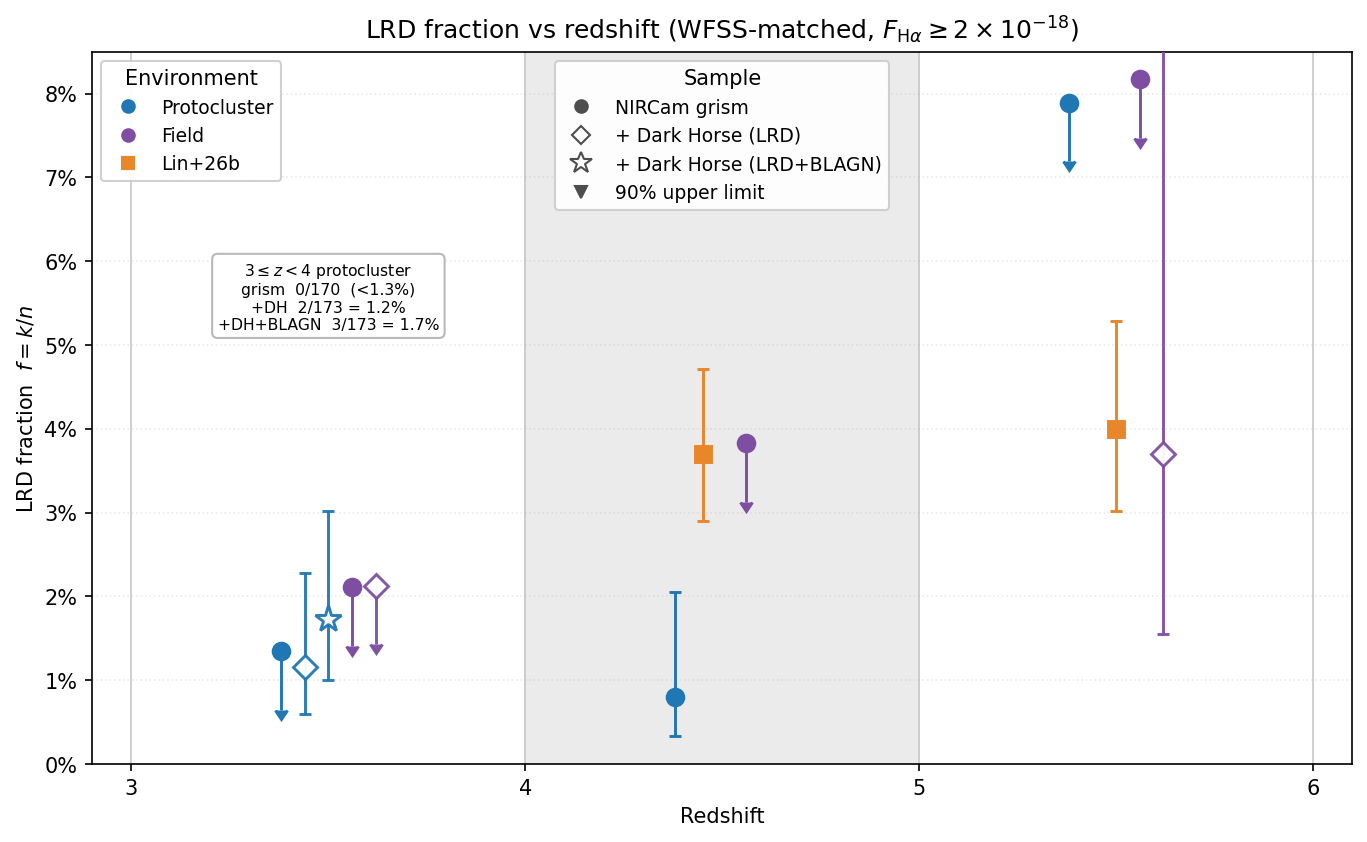}
    \caption{Fraction of LRDs as a function of redshift, matched to the WFSS H$\alpha$ depth. In each redshift bin $f=k/n$, where $n$ is the number of detectable HAEs (galaxies with a grism H$\alpha$ detection passing $F_{\mathrm{H}\alpha}\geq2\times10^{-18}\, \mathrm{erg\,s^{-1}\,cm^{-2}}$, enforced in $4\leq z<6$) and $k$ the LRDs among them. Points show the grism-only sample, the Dark Horse--extended sample (LRD only), and the combined LRD$+$BLAGN fraction; error bars are Jeffreys 68\% intervals, and downward arrows are one-sided 90\% upper limits for $k=0$. The LRD fractions in protoclusters (blue) and the field (purple) are shown separately, with the reconstructed AGN fractions of \citet{2026ApJ...997...61L} (orange) shown for comparison. In every bin, the protocluster and field fractions are consistent; at $3\leq z<4$, both are zero in the grism-only sample.}
    \label{fig:frac}
\end{figure*}

\subsection{LRD Fraction} \label{sec:fraction}

In each redshift bin, the LRD fraction $f=k/n$ is measured, where $n$ is the number of detectable HAEs and $k$ is the number classified as LRDs. The fractions are computed separately for the protocluster and field populations, as well as for the LRD-only and the combined LRD$+$BLAGN numerators (Table~\ref{tab:binom}). Figure~\ref{fig:frac} shows the results. In the grism-only sample, we find no protocluster LRDs at $3\le z<4$ ($0/170$; $<1.3\%$ at 90\% confidence). The Dark Horse extension and the combined LRD$+$BLAGN numerator recover a small nonzero value there ($\sim1.2\%$ and $\sim1.7\%$, respectively). At higher redshift, the fraction is $0.8\%$ at $4\le z<5$, with $0/28$ at $5\le z<6$. Across the full range, the LRD fraction in protoclusters is low ($\lesssim2\%$) with no significant redshift trend once luminosity-matched (Section~\ref{sec:completeness}). It lies below the $\sim3.9\%$ AGN fraction measured at  $3.9<z<6$ \citep{2026ApJ...997...61L}. Below, we show this offset is not driven by environment.

\begin{table*}
\centering
\caption{Protocluster LRD fractions compared with the field and with \citet{2026ApJ...997...61L}. In each redshift bin, $k/n$ is the number of LRDs (or LRD$+$BLAGN) over detectable HAEs. In parentheses, $p_{\rm Lin}$ is the one-sided binomial probability of observing $\le k$ sources if the true rate equalled the pooled \citet{2026ApJ...997...61L} AGN fraction (3.83\%; Appendix~\ref{app:lin}). Bold marks $p<0.05$. The Grism$+$DH (LRD$+$BLAGN) column is the fair comparison to their AGN fraction. The field contains no BLAGN, so one field column applies to both Dark Horse--extended numerators. $p_{\rm Fisher}$ is a two-sided Fisher exact test of protocluster versus field in the same bin. In the grism-only sample, the field contains no LRDs, and $p_{\rm Fisher}=1$ in every bin.}
\label{tab:binom}
\begin{tabular}{@{}lcccccc@{}}
\toprule
 & \multicolumn{3}{c}{Protocluster $k/n$ ($p_{\rm Lin}$)} & Field $k/n$ & \multicolumn{2}{c}{$p_{\rm Fisher}$} \\
\cmidrule(lr){2-4} \cmidrule(lr){6-7}
$z$ bin & Grism (LRD) & Grism$+$DH (LRD) & Grism$+$DH (LRD$+$BLAGN) & Grism$+$DH & LRD & LRD$+$BLAGN \\
\midrule
$2.5\le z<3$ & 0/14 (0.579)            & 0/14 (0.579)          & 0/14 (0.579)          & 0/15  & 1.00 & 1.00 \\
$3\le z<4$   & 0/170 (\textbf{0.0013}) & 2/173 (\textbf{0.037}) & 3/173 (0.099)        & 0/107 & 0.53 & 0.29 \\
$4\le z<5$   & 1/125 (\textbf{0.046})  & 1/125 (\textbf{0.046}) & 1/125 (\textbf{0.046}) & 0/59 & 1.00 & 1.00 \\
$5\le z<6$   & 0/28 (0.336)            & 0/28 (0.336)          & 0/28 (0.336)          & 1/27  & 0.49 & 0.49 \\
\midrule
All          & 1/337                   & 3/340                 & 4/340                 & 1/208 & 1.00 & 0.65 \\
\bottomrule
\end{tabular}
\end{table*}

\subsection{Environment Independence} \label{sec:environment}

To separate environmental effects from redshift evolution, we compared the protocluster and field LRD fractions within each redshift bin. In every bin, the two are statistically indistinguishable. A two-sided Fisher exact test of the protocluster and field counts gives $p\ge0.29$ in every bin (Table~\ref{tab:binom}) and $p=0.65$ pooled over the four redshift bins. The largest difference is at $3\le z<4$, where the field contains no LRD$+$BLAGN sources (0/107) against 3/173 in protoclusters, and it is not significant. A logistic regression of LRD membership on redshift and environment likewise yields no significant dependence on either (environment coefficient $p=0.55$; redshift $p=0.39$; protocluster and field, flux-matched). However, these tests are limited by a small number of LRDs. We find no evidence that the LRD fraction depends on large-scale environment at $3\le z<6$. This is consistent with the environment-independent AGN fraction reported by \citet{2026ApJ...997...61L} at $3.9<z<6$. This null is also robust to the clustering parameters (Appendix~\ref{app:clustering}).  Varying the minimum membership from 15 to 25 changes the number of candidates from 26 to 12, and varying either linking length by $\pm25\%$ changes it by at most one. No setting yields a significant environmental difference (two-sided Fisher $p\ge0.38$). The null also holds when only the 12 candidates that are significant overdensities are retained.

\begin{figure*}[htbp]
    \centering
    \includegraphics[width=\textwidth]{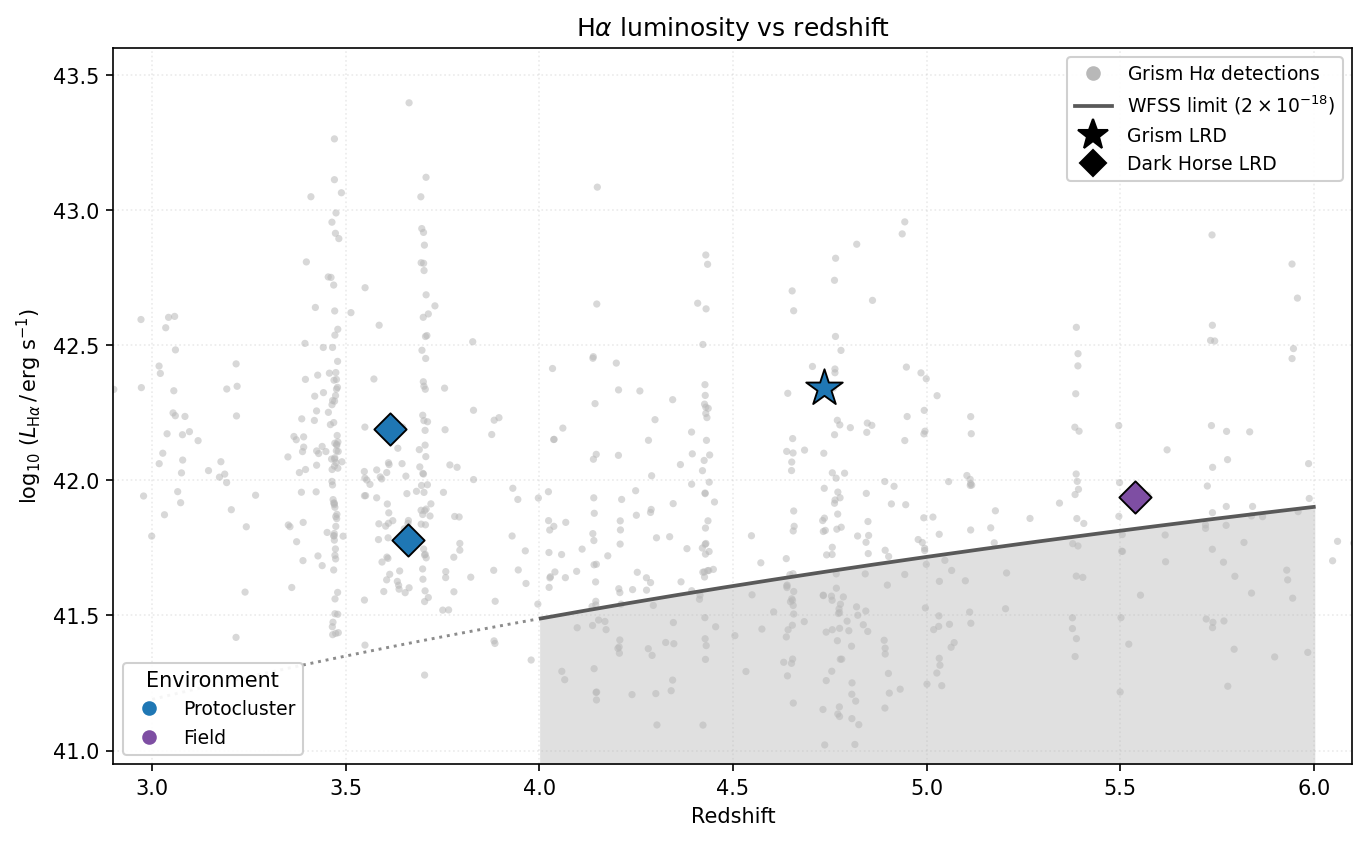}
    \caption{H$\alpha$ luminosity versus redshift for the LRD candidates with a grism H$\alpha$ detection (colored by environment; star: grism-selected, diamond: Dark Horse), over the full grism-H$\alpha$-detected population (gray) for context. The curve marks the luminosity corresponding to the WFSS flux limit $F_{\mathrm{H}\alpha}=2\times10^{-18}\,\mathrm{erg\,s^{-1}\,cm^{-2}}$, separating detectable (above) from undetectable (below) sources; it is drawn solid where the cut is enforced ($4\leq z<6$, with the excluded region shaded) and dotted as a reference at $z<4$. LRDs lacking a grism H$\alpha$ flux (the NIRSpec-only Dark Horse sources) are omitted.}
    \label{fig:lum}
\end{figure*}

\subsection{Completeness} \label{sec:completeness}

We observe that the protocluster fractions lie a factor of a few below the GOODS-N values of \citet{2026ApJ...997...61L}. Compared naively---our flux-limited, LRD-only counts against their AGN fraction---this offset appears significant at $3\le z<4$ ($p_{\rm Lin}=1.3\times10^{-3}$, $\approx3\sigma$; Table~\ref{tab:binom}). This offset shrinks as the selections are matched. It remains nominally significant after adding the Dark Horse LRDs ($p_{\rm Lin}=0.037$), but not once the BLAGN are included to match their broad-line plus V-shaped definition  ($p_{\rm Lin}=0.099$). Because a fixed flux limit corresponds to a luminosity floor that rises with redshift, we tested whether this offset is a selection effect by recomputing the fractions under the fixed luminosity cut $L_{\mathrm{H}\alpha}>10^{41.5}\,\mathrm{erg\,s^{-1}}$ (Appendix~\ref{app:lin}, Table~\ref{tab:lummatch}) used by \citet{2026ApJ...997...61L}. The combined protocluster fraction is then $1.8\%$ ($3/167$) at $3\le z<4$ and $0.65\%$ ($1/154$) at $4\le z<5$; the $3\le z<4$ value is not significantly below the literature rate ($p=0.11$). Only the $4\le z<5$ bin is marginally below ($p=0.02$), and it rests on a single source. Figure~\ref{fig:lum} shows that the LRDs sit above the luminosity floor, confirming that the low fractions are not caused by our sensitivity. The residual offset from \citet{2026ApJ...997...61L} is a cross-field comparison (our GOODS-S versus their GOODS-N). Because our field population shows the same shortfall, we attribute the residual offset to cosmic variance and selection differences rather than environment. In particular, our sample is not complete to the V-shaped-SED sources that \citet{2026ApJ...997...61L} include. Hence, our fractions are conservative lower bounds relative to their selection.

Under the same luminosity cut, the protocluster fraction shows no significant redshift trend. A logistic regression of LRD membership on redshift gives $p=0.58$ ($p=0.32$ for LRD$+$BLAGN) with a negative best-fit slope. A permutation test that does not rely on large-sample approximations gives consistent results (Appendix~\ref{app:lin}).

In addition, our environmental classification is incomplete. A protocluster that falls below the DBSCAN membership threshold, for example because it is sparsely sampled by the spectroscopy, is classified as field. LRDs in undetected protoclusters would thus dilute the contrast between the two populations and bias the comparison toward a null. Of the three field LRDs, only ID~5070 lies near this threshold (16 neighbors within the linking volume, against the 20 required); it joins a protocluster under more permissive settings, which leave the null unchanged (Table~\ref{tab:clusterrobust}). IDs~634042 and 642396 have 10 and 3 neighbors. Because the sampling density of the spectroscopic catalog declines with redshift, protocluster detection is also less complete at $z\gtrsim5$.

\section{Discussion} \label{sec:discussion}

\subsection{Environment-Independent LRD Fraction} \label{sec:disc_env}

We find that the LRD fraction among HAEs shows no detectable dependence on large-scale environment across $3\le z<6$, extending the environment-independent fraction of \citet{2026ApJ...997...61L} to $3\le z<4$. The two results agree despite different environmental measures, namely a continuous HAE overdensity in  GOODS-N and discrete protoclusters in GOODS-S. Our result also mirrors the weak clustering of low-luminosity AGN relative to star-forming galaxies \citep{2026ApJ...997...61L}. Together, these results suggest that structures on $\gtrsim10$ cMpc scales do not strongly regulate the fraction of LRDs, and by extension the growth of their black holes. LRD activity is more plausibly regulated by processes internal to galaxies or by galaxy-scale interactions than by the large-scale cosmic structure.

\subsection{Implications for LRD Evolution} \label{sec:disc_evol}

Our measurement also bears on whether the decline in the LRD population below $z\sim4$ is physical or a selection effect \citep{2026arXiv260630253L,
2026arXiv260407138R}. By comparing LRDs to the HAE population at matched luminosity and redshift, our fraction is less sensitive to the depth and selection biases that complicate raw measurements of number density. Once luminosity-matched, the protocluster LRD fraction shows no significant redshift trend over $3\le z<6$; if anything, it rises toward lower redshift (Section~\ref{sec:completeness}).  This supports the view that at least part of the apparent low-redshift decline reflects selection rather than intrinsic evolution \citep{2026arXiv260630253L,
2026arXiv260407138R}. 

Returning to the questions posed in Section~\ref{sec:introduction}: on environment, the protocluster fraction at $z<3.9$ is indistinguishable from the field.  On evolution, we detect no significant decline in either the LRD-only or the combined LRD$+$BLAGN fraction, so our data do not require LRDs to transform into BLAGN or to fade. Because our sample is largely insensitive to compact, blue broad-line AGN, however, we cannot test whether LRDs evolve into such sources. Further, because low-redshift protoclusters are the descendants of high-redshift ones, an environment-independent fraction implies that LRDs found in $z\gtrsim4$ protoclusters do not leave a distinctive over- or under-abundance of active black holes in more evolved descendant protoclusters at $3\le z<4$. Whatever their eventual fate, the LRD fraction tracks the field.

\subsection{Caveats and Future Work} \label{sec:disc_caveats}

Our conclusions are limited by small-number statistics. With each redshift bin having only single-digit counts of LRDs, our null result on the environment---that the LRD fractions in the protoclusters and field are statistically indistinguishable---rules out large environmental differences. However, it does not rule out modest ones. As discussed in Section~\ref{sec:completeness}, the comparison to \citet{2026ApJ...997...61L} is between different fields (GOODS-S versus GOODS-N), and our fractions are conservative lower bounds relative to theirs. Larger, uniformly selected samples spanning multiple fields \citep[e.g., NEXUS;][]{2026ApJ...999...31Z} will be needed to sharpen the environmental constraint and to establish whether the LRD fraction truly evolves below $z\sim4$.

\section{Summary} \label{sec:summary}
We measured the fraction of LRDs among HAEs in JADES protoclusters and field environments over $3 \le z < 6$, using GTO-4540 NIRCam grism spectroscopy extended by the Dark Horse NIRSpec classifications. Our main findings are:

\begin{itemize}[itemsep=2pt, topsep=2pt, parsep=0pt]
    \item \textit{Protoclusters (Figure~\ref{fig:hosts}).}  Density-based clustering yields 16 protocluster candidates. Five host at least one LRD or BLAGN. Their velocity dispersions ($\sigma_v\sim870$--$1960\,\mathrm{km\,s^{-1}}$) and masses are order-of-magnitude proxies instead of virialized quantities.
    \item \textit{Sample (Figure~\ref{fig:fits}).} We identify eight LRDs at $3.4\le z\le6.6$ and two BLAGN. The grism yields one broad-line LRD (ID~188191). The remaining LRDs rest on Dark Horse classifications, and only ID~188191 lies near the broad-line threshold. Our environmental comparisons hold with or without it.
    \item \textit{No detected environmental dependence (Figure~\ref{fig:frac}).} The protocluster LRD fraction is low ($\lesssim2\%$). In every redshift bin, the protocluster and field fractions are statistically indistinguishable. This extends the environment-independent AGN fraction of \citet{2026ApJ...997...61L} to $3\le z<4$.
    \item \textit{Selection, not environmental, offset (Figure~\ref{fig:lum}, Table~\ref{tab:lummatch}).} Our fractions lie below the \citet{2026ApJ...997...61L} GOODS-N rate. This is not a sensitivity effect since the LRDs sit above our flux limit. Because our field population shows the same shortfall relative to \citet{2026ApJ...997...61L}, we attribute the offset to cosmic variance and selection differences rather than the environment. Once luminosity-matched, the $3\le z<4$ fraction is consistent with the literature ($p=0.11$).
    \item \textit{No significant redshift trend.} Once luminosity-matched, the protocluster fraction shows no significant change over $3\le z<6$. This is consistent with at least part of the $z\sim4$ decline in LRD number density being a selection effect.
\end{itemize}

The environment-independent, field-like fraction of LRDs suggests that LRD black-hole growth is not strongly regulated by $\gtrsim10$ cMpc structure. Larger multi-field samples will be needed to improve the constraint and test for evolution below $z\sim4$.

\begin{acknowledgments}
This work is based on observations made with the NASA/ESA/CSA James Webb Space Telescope. The data were obtained from the Mikulski Archive for Space Telescopes at the Space Telescope Science Institute, which is operated by the Association of Universities for Research in Astronomy, Inc., under NASA contract NAS 5-03127 for JWST. These observations are associated with programs \#4540 and \#3215. We used the JADES NIRCam photometric catalog and imaging of the JOF (v2.0; \citealt{2025ApJS..281...50E}). The JADES photometric catalog is available at MAST via \dataset[doi:10.17909/z2gw-mk31]{https://doi.org/10.17909/z2gw-mk31}. Image cutouts were obtained from the GTO-4540 visual-inspection products.

This work began as a junior thesis for the course \textit{Astronomy 98: Research Tutorial in Astrophysics} at Harvard University. CAG thanks Morgan MacLeod for helpful discussions throughout the course. Subsequent development of the project was supported by the Harvard College Research Program (HCRP).

Support for program \#3215 was provided by NASA through a grant from the Space Telescope Science Institute, which is operated by the Association of Universities for Research in Astronomy, Inc., under NASA contract NAS5-03127. DJE is further supported as a Simons Foundation Investigator and by JWST/NIRCam contract to the University of Arizona, NAS5-02105.   BER acknowledges support as a Distinguished Visitor at the Institute of Astronomy at the University of Cambridge. AJB acknowledges funding from the ``FirstGalaxies'' Advanced Grant from the European Research Council (ERC) under the European Union's Horizon 2020 research and innovation program (Grant agreement No. 789056).
\end{acknowledgments}

\facilities{JWST (NIRCam, NIRSpec)}

\software{astropy \citep{2013A&A...558A..33A, 2018AJ....156..123A, 2022ApJ...935..167A}, NumPy \citep{2020Natur.585..357H}, SciPy \citep{2020NaMet..17..261V},  scikit-learn \citep{2011JMLR...12.2825P},  statsmodels \citep{seabold2010statsmodels}, Matplotlib \citep{2007CSE.....9...90H}, FITSmap \citep{2022A&C....3900586H}}

\section*{AI Disclosure}
 The authors declare the use of generative AI in the research and writing process. According to the GAIDeT taxonomy \citep{Suchikova03042026}, the following tasks were delegated to GAI tools under full human supervision:
\begin{itemize}[nosep]
    \item code generation and optimization;
    \item visualization;
    \item reformatting;
    \item proofreading and editing.
\end{itemize}
The GAI tools used were GitHub Copilot in Visual Studio Code (Claude Sonnet 4 and 4.5, Claude Haiku 4.5, and GPT-5) and Claude Code (Opus 4.8 and 5.5). All AI-assisted text and code were reviewed and verified by the authors. Responsibility for the final manuscript lies entirely with the authors. GAI tools are not listed as authors and do not bear responsibility for the final outcomes.
\bibliographystyle{aasjournalv7.1}
\bibliography{main,non_ads}

\appendix

\section{Protocluster Clustering} \label{app:clustering}
About the sample-median center $(\alpha_0,\delta_0)$, we assign each galaxy the transverse comoving offsets  $x=\chi(z)(\alpha-\alpha_0)\cos\delta_0$, $y=\chi(z)(\delta-\delta_0)$ and the line-of-sight distance $\chi(z)$. We scale the two transverse offsets by $\ell_\perp=10$ cMpc and the line-of-sight distance by $\ell_\parallel=20$ cMpc. Then, we run DBSCAN with a unit search radius and a minimum of 20 members \citep{10.5555/3001460.3001507}, following \citet{2015MNRAS.452.2528M}. The anisotropy reflects redshift-space distortion  \citep{1987MNRAS.227....1K} and the scales expected for protoclusters \citep{2017ApJ...844L..23C}.

We tested the sensitivity of the clustering by varying each hyperparameter about the fiducial $(\ell_\perp,\ell_\parallel,\mathrm{min\_samples})=(10\,\mathrm{cMpc},20\,\mathrm{cMpc},20)$, holding the other two fixed  (Table~\ref{tab:clusterrobust}). The number of recovered protocluster candidates is stable to the linking lengths: a $\pm25\%$ change in $\ell_\perp$ or $\ell_\parallel$ moves this number by at most one. However, it is more sensitive to minimum membership (e.g., 26 candidates at $\mathrm{min\_samples}=15$, but 12 candidates at $\mathrm{min\_samples}=25$). This is expected since the minimum membership parameter sets the density threshold directly. The LRD protocluster/field assignments separate into robust and boundary cases: three LRDs (5756, 160128, 188191) are recovered as protocluster members and two (634042, 642396) as field members under every setting tested. In contrast, three sources at $z\approx3.4$--$3.7$ (13064, 35453, 5070) migrate between field and protocluster as the linking volume changes, as expected for galaxies near a group boundary in the crowded low-redshift volume.

\begin{table*}[htbp]
\caption{Robustness of the clustering and of the environmental null to the DBSCAN parameters. Each DBSCAN parameter $(\ell_\perp, \ell_\parallel$, or the minimum membership min\_samples) is varied about the fiducial value (top row), holding the other two fixed. $N_{\rm cand}$ is the number of protocluster candidates (identical with and without the added Dark Horse sources); ``LRD p/f'' is the protocluster/field split of the eight LRDs, with the LRDs that change assignment relative to the fiducial listed next ($\to$f: to field; $\to$p: to protocluster). $f_{\rm proto}$ and $f_{\rm field}$ are the combined LRD$+$BLAGN fractions pooled over the four bins (fiducial: $4/340$ and $1/208$); $p_{\rm Fisher}$ is a two-sided Fisher exact test of protocluster versus field, and $p_{\rm env}$ is the $P>|z|$ of the environment coefficient in the logistic regression $P(\mathrm{LRD})\sim z+\mathrm{environment}$.}
\label{tab:clusterrobust}
\centering
\begin{tabular}{@{}llcclcccc@{}}
\toprule
Parameter & Value & $N_{\rm cand}$ & LRD p/f & Migrating LRDs &
$f_{\rm proto}$ (\%) & $f_{\rm field}$ (\%) & $p_{\rm Fisher}$ & $p_{\rm env}$ \\
\midrule
\multicolumn{2}{@{}l}{\textbf{Fiducial}: $(\ell_\perp,\ell_\parallel,\mathrm{min\_samples})=(10,20,20)$}
  & \textbf{16} & \textbf{5/3} & --- & \textbf{1.2} & \textbf{0.5} & \textbf{0.65} & \textbf{0.55} \\
\midrule
$\ell_\perp$ (cMpc)     & 7.5  & 16 & 4/4 & 13064\,$\to$f        & 1.2 & 0.4 & 0.41 & 0.45 \\
                        & 12.5 & 17 & 6/2 & 5070\,$\to$p         & 1.1 & 0.6 & 1.00 & 0.67 \\
\midrule
$\ell_\parallel$ (cMpc) & 15   & 15 & 3/5 & 13064, 35453\,$\to$f & 1.0 & 0.8 & 1.00 & 0.96 \\
                        & 25   & 17 & 6/2 & 5070\,$\to$p         & 1.0 & 0.7 & 1.00 & 0.86 \\
\midrule
$\mathrm{min\_samples}$ & 15   & 26 & 6/2 & 5070\,$\to$p         & 0.9 & 1.0 & 1.00 & 0.84 \\
                        & 25   & 12 & 4/4 & 13064\,$\to$f        & 1.3 & 0.4 & 0.38 & 0.31 \\
\bottomrule
\end{tabular}
\end{table*}

\begin{table}[htbp]
\centering
\caption{The 16 protocluster candidates in the combined grism$+$Dark Horse catalog, ordered by mean redshift $\bar z$. $N$ is the number of members, and $\sigma_v=c\,\sigma_z/(1+\bar z)$ is a line-of-sight dispersion proxy, where $\sigma_z$ is the standard deviation of the members' redshifts. $\delta=N/N_{\rm exp}-1$ is the overdensity relative to the number expected $N_{\rm exp}$ in the candidate's redshift range from the redshift distribution smoothed with a Gaussian kernel of width $0.1$ in redshift. $P$ is the Poisson probability of at least $N$ galaxies given $N_{\rm exp}$. Letters mark the five hosts shown in Figure~\ref{fig:hosts}; host B comprises two redshift peaks (Section~\ref{sec:samples}). Candidates at $z<2.5$ lie below our H$\alpha$ redshift range and do not enter the fractions.}
\label{tab:cands}
\begin{tabular}{@{}cccrrlc@{}}
\toprule
$\bar z$ & $z$ range & $N$ & $\sigma_v$ (km\,s$^{-1}$) & $\delta$ & $P$ & Host \\
\midrule
1.096 & 1.078--1.112 & 60 & 1200 &  2.67 & $9\times10^{-17}$ & \\
1.223 & 1.218--1.235 & 22 &  650 &  2.21 & $3\times10^{-6}$  & \\
1.544 & 1.537--1.554 & 22 &  660 &  4.67 & $2\times10^{-10}$ & \\
2.559 & 2.538--2.578 & 29 & 1060 &  1.45 & $2\times10^{-5}$  & E \\
2.686 & 2.656--2.703 & 45 &  990 &  1.58 & $3\times10^{-8}$  & \\
2.810 & 2.776--2.834 & 41 & 1020 &  1.08 & $2\times10^{-5}$  & \\
3.465 & 3.419--3.493 & 99 & 1140 &  1.38 & $3\times10^{-14}$ & A \\
3.600 & 3.562--3.638 & 34 & 1330 & $-0.16$ & 0.86            & D \\
3.700 & 3.660--3.730 & 59 &  870 &  0.81 & $2\times10^{-5}$  & C \\
4.145 & 4.110--4.177 & 27 &  730 &  0.62 & 0.012             & \\
4.426 & 4.381--4.455 & 46 &  830 &  1.30 & $5\times10^{-7}$  & \\
4.650 & 4.605--4.685 & 34 &  730 &  0.39 & 0.038             & \\
4.782 & 4.705--4.860 & 82 & 1960 &  0.35 & 0.006             & B \\
5.018 & 4.978--5.065 & 29 & 1280 &  0.51 & 0.023             & \\
5.386 & 5.343--5.406 & 22 &  570 &  1.91 & $2\times10^{-5}$  & \\
5.753 & 5.720--5.804 & 28 & 1140 &  1.29 & $8\times10^{-5}$  & \\
\bottomrule
\end{tabular}
\end{table}

At each setting, we also recomputed the detectable-HAE denominators, the combined (LRD$+$BLAGN) protocluster and field fractions, and the environmental tests. Both fractions remain low across the grid (protocluster 0.9--1.3\%, field 0.4--1.0\%). Further, neither a two-sided Fisher exact test ($p\ge0.38$) nor the environment coefficient of the logistic regression ($p_{\rm env}=0.31$--$0.96$) indicates a significant environmental dependence.

DBSCAN identifies a group wherever at least 20 galaxies fall within the linking volume. Because the JOF is about one $\ell_\perp$ wide, this volume spans nearly the whole field over a $\pm20$\,cMpc slice in redshift, so where the survey samples galaxies densely, the average density alone can approach the threshold. Indeed, when we smooth the catalog's redshifts to erase real redshift spikes while preserving the overall redshift distribution, DBSCAN still returns 5--10 candidates at $2.5\le z<6$. A detection therefore does not by itself establish an overdensity. For each candidate, we compare its number of members $N$ with the number $N_{\rm exp}$ expected in the same redshift range from the smoothed redshift distribution (Gaussian kernel of width $0.1$ in redshift, which erases spikes but follows the survey's overall selection), and compute $\delta=N/N_{\rm exp}-1$ and the Poisson probability $P$ of at least $N$ members (Table~\ref{tab:cands}). Twelve of the 16 candidates are significant overdensities ($P<0.01$; $\delta\approx0.35$--4.7). The four that are not lie at $\bar z=3.60$, 4.15, 4.65, and 5.02. Reassigning their members to the field leaves the environmental null unchanged (two-sided Fisher $p\ge0.44$ in every bin; $p=0.67$ pooled). Testing the method on mock catalogs from cosmological simulations, in which true protocluster membership is known, is left to future work.

\section{Emission-Line Fitting and Model Selection} \label{app:fitting}
We fit the continuum-subtracted H$\alpha$ window with four Gaussian models: (1) single H$\alpha$; (2) H$\alpha$ + [N\,{\sc ii}]\,$\lambda\lambda6548,6583$ with amplitude ratio $A_{6548}=A_{6583}/3$ \citep{2023AdSpR..71.1219D}; (3) narrow + broad H$\alpha$; (4) narrow + broad H$\alpha$ + [N\,{\sc ii}]. We select the model that minimizes $\mathrm{BIC}=\chi^2+k\ln n$, requiring $\Delta\mathrm{BIC}>5$ \citep{2007MNRAS.377L..74L,2026Natur.653.1017J}. Here, $k$ is the number of free parameters in the model, and $n$ is the number of resolution elements from the grism resolving power \citep{2017jwst.rept.5001G}.  A lower BIC indicates a model that better balances goodness-of-fit and model simplicity.  With BICs calculated, the selection logic is as follows:
 \begin{enumerate}
     \item Compare Model 1 vs 2 and Model 1 vs 3 separately.
     \item Choose whichever of \{2, 3\} significantly improves on Model 1. Otherwise, keep Model 1.
     \item If Model 2 or 3 was chosen, compare it to Model 4. Adopt Model 4 only if it significantly improves on 2 or 3. Otherwise, retain the simpler model.
 \end{enumerate}

Broad-line sources have $\mathrm{FWHM}=c\,\Delta\lambda/\lambda_{\rm center}>1000\,\mathrm{km\,s^{-1}}$ \citep{2001AJ....122..549V,2015ApJ...813...82R}. The best-fit parameters of the five sources with a grism fit are listed in Table~\ref{tab:fits}.

\begin{table*}[htbp]
\centering
\caption{Best-fit H$\alpha$ parameters for the sample sources with a NIRCam grism fit. $C$ is the continuum and $A$ the Gaussian peak amplitude (both in $\mu$Jy); FWHM is in km s$^{-1}$. For Model~1, the single Gaussian is listed under the narrow columns.}
\label{tab:fits}
\begin{tabular}{@{}lccccccccc@{}}
\toprule
ID & Type & Filter & Model & $C$ ($\mu$Jy) & $A_{\rm narrow}$ ($\mu$Jy) &
FWHM$_{\rm narrow}$ & $A_{\rm broad}$ ($\mu$Jy) & FWHM$_{\rm broad}$ & BIC \\
\midrule
188191 & LRD   & F322W2 & 3 & 0.31  & 7.12 & 179  & 2.08    & 1013    & 104.9 \\
160128 & LRD   & F322W2 & 1 & 0.10  & 5.91 & 577  & \nodata & \nodata & 97.4  \\
35453  & LRD   & F322W2 & 1 & 0.056 & 2.58 & 513  & \nodata & \nodata & 108.6 \\
634042 & LRD   & F444W  & 1 & 0.16  & 1.05 & 768  & \nodata & \nodata & 113.7 \\
171973 & BLAGN & F322W2 & 1 & 0.16  & 0.91 & 1501 & \nodata & \nodata & 67.8  \\
\bottomrule
\end{tabular}
\end{table*}

\section{Robustness of Broad-line and Compactness Criteria} \label{app:robustness}
All seven LRDs with a photometric counterpart have F444W CIRC3/CIRC1 $\le1.07$, satisfying the stricter \citet{2025ApJ...985..119Z} threshold of $1.2$ (Table~\ref{tab:compact}). The classification is thus insensitive to aperture choice. ID~642396 ($z=6.55$) lies within the v2.0 imaging but is absent from the v2.0 photometric catalog (the nearest source lies $2.2''$ away); it appears only in a later JADES catalog, and it lies outside our redshift bins.  ID~176606 (CIRC3/CIRC1 $=1.33$) is the only grism candidate resolved at this aperture and is excluded as an LRD (Section~\ref{sec:lrdsample}).

Broad/narrow classifications agree across the filters in which each source is fit, with one exception: ID~188191 is fit with a narrow+broad pair in both F322W2 and F356W, but the second component's FWHM straddles the threshold (1013 and 612 km s$^{-1}$ in F322W2 and F356W, respectively). We retained it as marginal, and our environmental comparisons hold without it. Another grism candidate, ID~175930, appears broad only in a BIC-disfavored filter and is excluded.

\begin{table}[htbp]
\centering
\caption{Compactness of the LRD candidates under two F444W aperture ratios. Every retained candidate with a photometric counterpart satisfies the stricter \citet{2025ApJ...985..119Z} threshold ($<1.2$) at the fiducial aperture ($0.25''/0.10''$) and remains compact at the tighter aperture ($0.15''/0.10''$). ID~642396 is not in the v2.0 photometric catalog (see text).}
\label{tab:compact}
\begin{tabular}{@{}lccc@{}}
\toprule
ID & CIRC3/CIRC1 & CIRC2/CIRC1 & Compact \\
   & ($0.25''/0.10''$) & ($0.15''/0.10''$) & \\
\midrule
35453  & 1.008 & 1.010 & Yes \\
634042 & 1.026 & 1.015 & Yes \\
5070   & 1.039 & 1.012 & Yes \\
160128 & 1.045 & 1.023 & Yes \\
13064  & 1.050 & 1.023 & Yes \\
5756   & 1.060 & 1.037 & Yes \\
188191 & 1.068 & 1.030 & Yes \\
642396 & \nodata & \nodata & \nodata \\
\bottomrule
\end{tabular}
\end{table}

\section{Comparison with Lin et al. (2026b)} \label{app:lin}
We reconstructed the HAE denominators of \citet{2026ApJ...997...61L} by back-solving their reported AGN fractions (3.7\% at $3.9<z<5$, 4.0\% at $5<z<6$) against the AGN counts in their Tables 1 and 2, which gives 28 AGN among 732 HAEs over $3.9<z<6$. These helped compute the matching Jeffreys credible intervals. To compare at fixed depth, we adopt their luminosity cut $L_{\mathrm{H}\alpha}>10^{41.5}\, \mathrm{erg\,s^{-1}}$. Our WFSS flux limit reaches this only at $z\lesssim4$, so at higher $z$, our sample is incomplete between $10^{41.5}\mathrm{erg\,s^{-1}}$ and our flux floor. The luminosity-matched protocluster fractions are given in Table~\ref{tab:lummatch}. Within this luminosity-matched protocluster sample, a logistic regression of LRD membership on redshift finds no significant trend ($p=0.58$ for LRDs, $0.32$ for LRD$+$BLAGN). As a check that does not rely on large-sample approximations used by the logistic regression, we compare the mean redshift of the LRDs with that of 100{,}000 random draws of the same number of HAEs from the parent sample. The LRDs are not at unusually low or high redshift ($p=0.61$ for LRDs and $0.32$ for LRD$+$BLAGN), consistent with the logistic regression.

\begin{table}[htbp]
\centering

\caption{Protocluster LRD fraction under the fixed luminosity cut $L_{\mathrm{H}\alpha}>10^{41.5}$ erg s$^{-1}$ of \citet{2026ApJ...997...61L}. Columns list the grism-only LRD count, the Dark Horse--extended LRD count, and the combined LRD$+$BLAGN count. Each count is reported over the number of HAEs with $L_{\mathrm{H}\alpha}>10^{41.5}$ (no flux cut); $f$ is the combined fraction and $p_{\rm Lin}$ the one-sided binomial probability against the Lin et al.\ rate (3.83\%). The $5\le z<6$ value is a 90\% upper limit.}
\label{tab:lummatch}
\begin{tabular}{@{}lccccc@{}}
\toprule
$z$ bin & Grism (LRD) & Grism$+$DH (LRD) & Grism$+$DH (LRD$+$BLAGN) & $f$ (combined) & $p_{\rm Lin}$ \\
\midrule
$3\le z<4$ & 0/164 & 2/167 & 3/167 & $1.8\%$   & 0.115 \\
$4\le z<5$ & 1/154 & 1/154 & 1/154 & $0.65\%$  & 0.018 \\
$5\le z<6$ & 0/41  & 0/41  & 0/41  & $<5.5\%$  & 0.202 \\
\bottomrule
\end{tabular}
\end{table}

\end{document}